\documentclass{jpp}
\usepackage{physics}
\usepackage{graphicx}
\usepackage{caption}
\usepackage{wrapfig}
\usepackage{epstopdf}
\usepackage[toc,page]{appendix}
\usepackage{listings}
\usepackage{enumerate}
\usepackage{subcaption}
\usepackage{color}
\usepackage{paralist}
\usepackage[font=small]{caption}
\usepackage{booktabs}
\usepackage{todonotes}
\usepackage{setspace}
\usepackage{mathtools, cuted}
\usepackage[T1]{fontenc}
\usepackage[utf8]{inputenc}

\newcommand{\ensav}[2]{\left<{#1}\right>_{\vb*{#2}}}
\newcommand{\figref}[1]{figure~\ref{fig:#1}}
\newcommand{\Figref}[1]{Figure~\ref{fig:#1}}

\shorttitle{Transition to subcritical turbulence in a tokamak plasma}
\shortauthor{F. van Wyk and others}

\title{Transition to subcritical turbulence in a tokamak plasma}

\author{
F. van Wyk\aff{1,2,3}\corresp{\email{ferdinand.vanwyk@physics.ox.ac.uk}},
E. G. Highcock\aff{1,4},
A. A. Schekochihin\aff{1,5},
C. M. Roach\aff{2},
A. R. Field\aff{2}
\and W. Dorland\aff{1,6}
}

\affiliation{
\aff{1}Rudolf Peierls Centre for Theoretical Physics, University of Oxford, Oxford OX1 3NP, UK
\aff{2}CCFE, Culham Science Centre, Abingdon OX14 3DB, UK
\aff{3}STFC Daresbury Laboratory, Daresbury WA4 4AD, UK
\aff{4}Chalmers University of Technology, Fysikg{\aa}rden 1, Gothenberg, Sweden
\aff{5}Merton College, Oxford OX1 4JD, UK
\aff{6}Department of Physics, University of Maryland, College Park, MD 20742-4111, USA
}

\begin{document}

\maketitle

\begin{abstract}
  Tokamak turbulence, driven by the ion-temperature gradient and occurring in
  the presence of flow shear, is investigated by means of local, ion-scale,
  electrostatic gyrokinetic simulations (with both kinetic ions and electrons)
  of the conditions in the outer core of the Mega-Ampere Spherical Tokamak
  (MAST). A parameter scan in the local values of the ion-temperature gradient
  and flow shear is performed. It is demonstrated that the experimentally
  observed state is near the stability threshold and that this stability
  threshold is nonlinear: sheared turbulence is subcritical, i.e., the system
  is formally stable to small perturbations, but, given a large enough initial
  perturbation, it transitions to a turbulent state. A scenario for such a
  transition is proposed and supported by numerical results: close to
  threshold, the nonlinear saturated state and the associated anomalous heat
  transport are dominated by long-lived coherent structures, which drift across
  the domain, have finite amplitudes, but are not volume-filling; as the system
  is taken away from the threshold into the more unstable regime, the number of
  these structures increases until they overlap and a more conventional chaotic
  state emerges. Whereas this appears to represent a new scenario for
  transition to turbulence in tokamak plasmas, it is reminiscent of the
  behaviour of other subcritically turbulent systems, e.g., pipe flows and
  Keplerian magnetorotational accretion flows.
\end{abstract}

\section{Introduction}
  Controlling turbulence in magnetically confined plasmas is the key to
  achieving sustained nuclear fusion as an energy
  source~\citep{Krushelnick2005}. Typically, unstable perturbations driven by
  the pressure gradient and other sources of free energy grow exponentially and
  eventually saturate nonlinearly, leading to turbulence. Recent
  work~\citep{Newton2010, Highcock2010, Highcock2011, Barnes2011a,
  Schekochihin2012a} has shown that in the presence of sheared flows, such
  systems can be subcritical.  This means that all perturbations are linearly
  stable and a transition to a turbulent state only occurs if large enough
  initial perturbations undergo sufficient transient growth to allow nonlinear
  interaction. Understanding the transition to a turbulent state is a
  long-standing challenge in fluids~\citep{Barkley2015} and, more recently, in
  fusion plasmas, where a quiescent state leads to dramatically improved
  confinement. Experimental studies in simple devices~\citep{Weixing1993,
  Klinger1997, Riccardi1997, Burin2005} have proposed that this transition
  occurs through an increasing number of unstable frequencies leading to a
  turbulent state with a broadband spectrum. There is, however, currently very
  little known about a subcritical transition to turbulence in fusion-relevant
  plasmas. Here we use first-principles gyrokinetic simulations of a turbulent
  plasma in the outer core of the Mega-Ampere Spherical Tokamak (MAST) to
  demonstrate that the experimentally observed state is near the transition
  threshold, that the turbulence in this state is subcritical, and that
  transition to turbulence occurs via accumulation of long-lived, intense,
  finite-amplitude coherent structures, which dominate the near-threshold
  state. This represents a conceptually new and thus far unexplored scenario
  for transition to turbulence in magnetised plasmas.

  The seemingly abrupt transition to turbulence from a quiescent state is a
  fluid-dynamical phenomenon that has fascinated scientists since the seminal
  experiments by~\cite{Reynolds1883}. In fusion
  plasmas, transition to turbulence is a similarly tantalising challenge, both
  as a matter of fundamental physics and because understanding and controlling
  turbulence remains the greatest challenge to technologically and commercially
  effective fusion-power generation~\citep{Krushelnick2005}.  This challenge
  arises from the turbulent eddies' propensity for transporting heat and
  particles out of the core of the device, often leading to dramatically
  degraded plasma confinement~\citep{Kotschenreuther1995a}.

  Extensive experimental~\citep{Baker2001, Mantica2009, Ghim2014,Field2011a} and
  numerical~\citep{Kotschenreuther1995a,Dimits2000,Barnes2011a,Highcock2012,Field2011a,Roach2009}
  work has identified the key parameters that trigger the transition to a
  turbulent regime at certain critical values. In particular, the ion
  temperature gradient (ITG), $\kappa_T = -\dv*{\ln T_i}{r}$ ($T_i$ is the ion
  temperature, and $r$ is an appropriate dimensionless radial coordinate
  defined later), acts as a source of free energy~\citep{Coppi1967}, driving
  turbulent fluctuations, whereas differential toroidal rotation of the plasma,
  quantified by the (non-dimensionalised) flow shear perpendicular to the
  confining magnetic field, $\gamma_E = (r/q)\dv*{\omega}{r} (a/v_{{\rm th}i})$
  ($q$ is the ``magnetic safety factor'', $\omega$ is the angular frequency of
  toroidal rotation, $a$ is the minor radius of the toroidal device, and
  $v_{{\rm th}i} = \sqrt{2 T_i/m_i}$ is the ion thermal velocity) can suppress
  turbulence~\citep{Burrell1997a}.

  In a steady state, a given amount of power injected (or, in a fusion power
  plant, fusion-generated) in the core of the device must be transported out
  through the plasma. Since the heat flux typically increases with the
  temperature gradient, more power requires maintaining a larger ITG.\@ In the
  most straightforward scenario, above a certain threshold value of ITG, the
  plasma becomes linearly unstable, perturbations to the equilibrium are
  amplified and saturate, giving rise to a turbulent state.  In this state,
  heat is transported by turbulent fluctuations, whose amplitude becomes larger
  at larger ITG~\citep{Barnes2011}. As a result, heat flux typically scales very
  strongly with the ITG, leading to ``stiff transport'', with the practical
  consequence that the system cannot stray too far above the threshold value
  of ITG~\citep{Kotschenreuther1995a}.

  The situation becomes more complicated in the presence of differential
  rotation. Perpendicular flow shear has been shown to have a suppressing
  effect on the linear instabilities and can even render the plasma completely
  linearly stable, i.e., all modes decay exponentially at large times. However,
  this may still entail substantial transient growth of
  perturbations~\citep{Newton2010,Schekochihin2012a,Roach2009} and, given finite
  initial perturbations, can lead to a saturated nonlinear state --- a
  phenomenon known as ``subcritical''
  turbulence~\citep{Highcock2010,Highcock2011,Barnes2011a}. It is then an intriguing question
  how the heat flux due to such turbulence can change continuously from zero
  below the transition threshold to small but finite ITG-dependent values just above it (as
  it indeed does, in our simulations). In the case of supercritical turbulence,
  the saturated fluctuation amplitude everywhere increases continuously from
  small values near the threshold to finite ones far above it. In contrast, in
  the regime leading to subcritical turbulence, we find that
  small-amplitude perturbations decay and only finite-amplitude ones can
  survive and saturate. Therefore, the turbulent heat flux must increase with
  increasing ITG by some mechanism other than a continuous increase in
  fluctuation amplitude. Here we will identify this mechanism (which will lead
  us to a very different transition scenario than the conventional one outlined
  above) and ascertain that it is relevant to real experimental situations.

\section{Gyrokinetic simulations}
  As an example of such a real experimental situation, we consider turbulence
  in the MAST tokamak, which is a major current experimental machine that is
  well diagnosed and actively used to test novel fusion concepts.  We pick a
  magnetic configuration and plasma parameters describing the outer core of
  MAST for a particular discharge (\#27268; see~\citealt{Field2014a} for its
  detailed description).  We then solve numerically for the turbulent
  fluctuations in a local ``flux tube'', by means of
  gyrokinetic simulations~\citep{Dimits2000,Fasoli2016} with the widely used
  code GS2\footnote{http://gyrokinetics.sourceforge.net}. Our simulations are
  electrostatic, restricted to ion scales, include both kinetic ions and
  kinetic electrons, and model collisions using a linearised Fokker-Planck
  collision operator (see appendix~\ref{App:gk_numerics} for the specific
  equilibrium and resolution parameters used).
  \begin{figure}
    \centering
    \begin{subfigure}[t]{0.5\textwidth}
      \includegraphics[width=\textwidth]{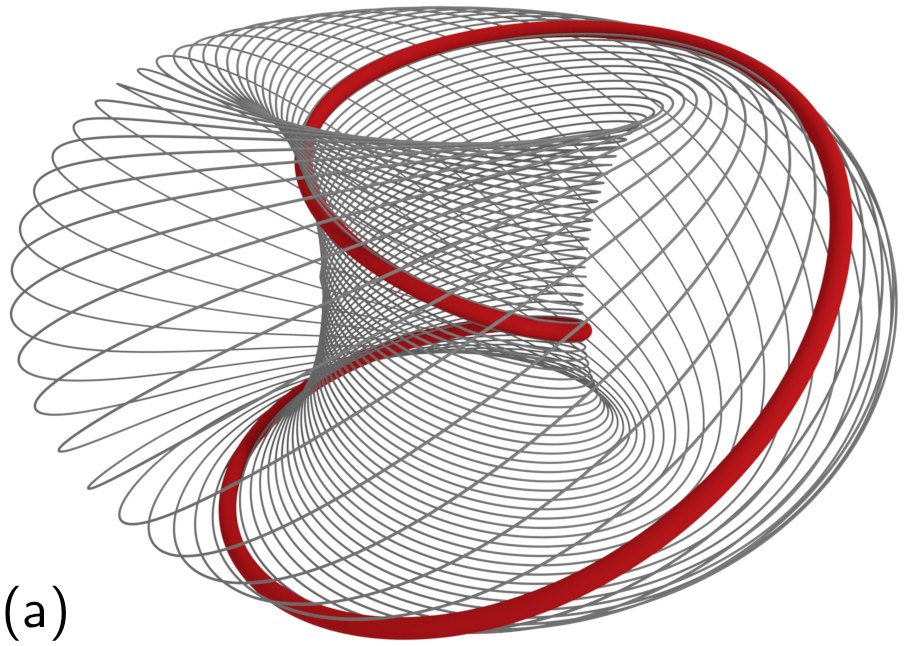}
      \caption{}
      \label{fig:mast}
    \end{subfigure}
    \hspace{3em}
    \begin{subfigure}[t]{0.3\textwidth}
      \includegraphics[width=\textwidth]{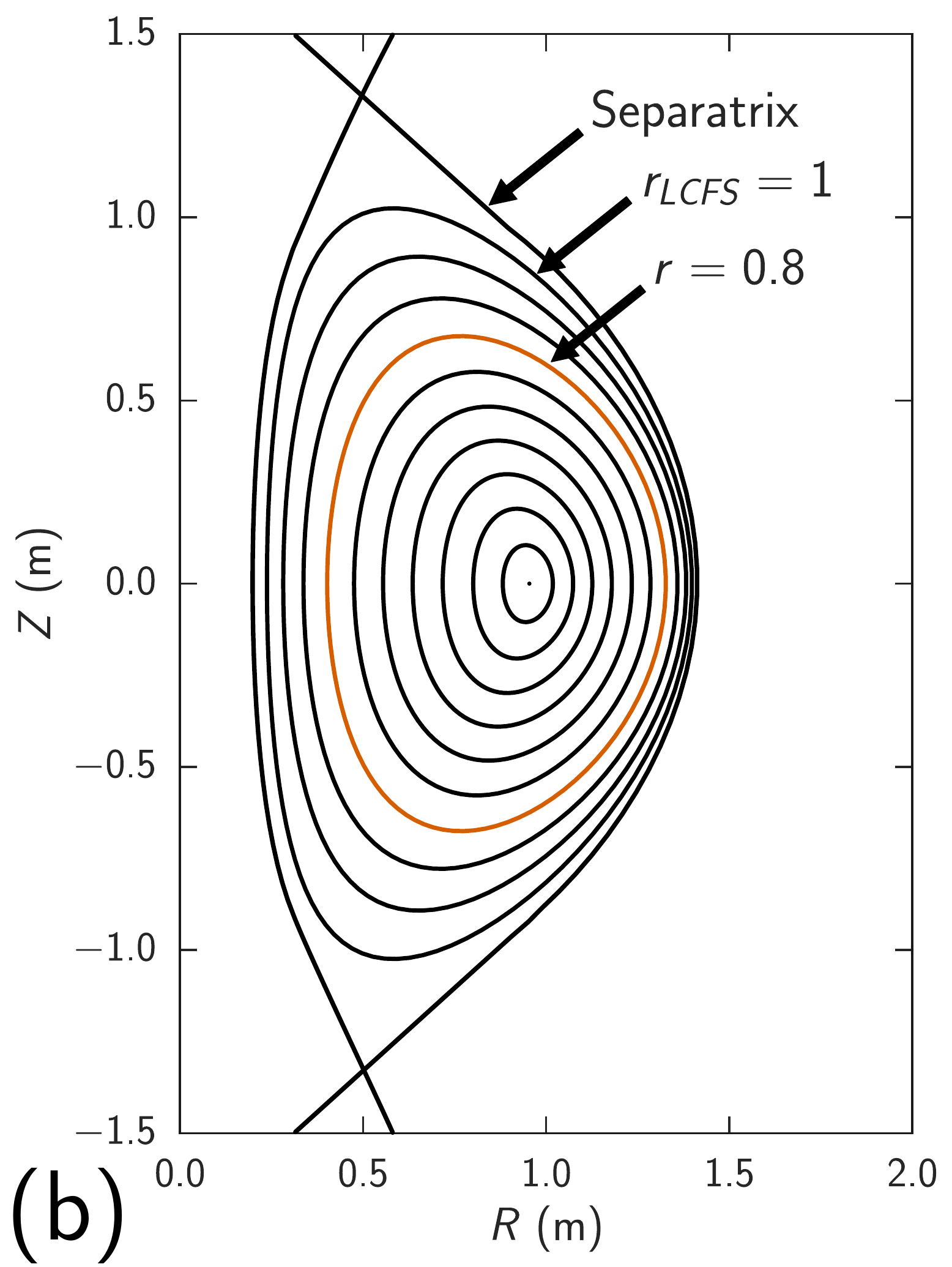}
      \caption{}
      \label{fig:nestedflux}
    \end{subfigure}
    \caption{(a) The flux surface at $r = 0.8$ traced by the field
      lines that lie in it. The field line marked in red is the centre line of
      a GS2 ``flux tube''. The actual GS2 flux tube is rectangular in the
      outboard midplane but, as the magnetic field is sheared, the flux tube
      twists as it follows the field line along the flux surface.
      (b) Poloidal projection of the MAST flux surfaces.
      The flux surface at $r = 0.8$, the last closed flux surface (LCFS) and
      the separatrix are indicated.  The~\cite{Miller1998b} radial
      coordinate $r$ of a flux surface is the ratio of the diameter of this
      flux surface at the elevation of the magnetic axis ($Z=0$ in this plot)
      to that of the last closed flux surface.
    }
  \end{figure}

  A GS2 flux tube is a twisted box of finite radial and poloidal width that
  follows a magnetic field line once around the tokamak in the poloidal
  direction (\figref{mast}). We assume that the underlying equilibrium is
  axisymmetric, so a single flux tube can be used to simulate the entire flux
  surface, allowing a large saving in computational cost. In this local
  approach, all equilibrium gradients (of density, velocity, and temperature)
  are assumed constant across the radial extent of the flux tube. We pick the
  time $t = 0.25$s from the beginning of the discharge and the radial location
  $r = D/2a = 0.8$, where $D$ is the diameter of the flux surface hosting our
  flux tube, at the height of the magnetic axis, and $2a$ is the diameter of
  the last closed flux surface (the ``edge'' of the plasma). This definition of
  the radial location is convenient because it coincides with
  the~\citet{Miller1998b} parametrisation of the flux tube geometry used by
  GS2. \Figref{nestedflux} shows a poloidal projection of the MAST flux
  surfaces for discharge \#27268 with the flux surface at $r = 0.8$
  highlighted.

  The gyrokinetic equation (\citealt{Frieman1982}; see
  appendix~\ref{App:gk_numerics}) solved by GS2 gives us the time evolution of
  the perturbed distribution function $\delta f_s(t, \vb*{r}, \vb*{v})$ of
  particles of species $s$ (ions or electrons), where $\vb*{r}$ is the spatial
  position and $\vb*{v}$ the velocity. Gyrokinetic theory makes use of the fact
  that turbulent fluctuations in a tokamak plasma occur at much longer time
  scales than the Larmor motion of the particles (although still much shorter
  than the evolution of the background thermal and magnetic equilibrium state).
  This allows the Larmor motion to be averaged over analytically, leading to a
  kinetics of Larmor rings, whose distribution depends on two, rather than
  three velocity variables (parallel $v_\parallel$ and perpendicular $v_\perp$
  velocities, but not the gyroangle), reducing the phase space from six to five
  dimensions --- a more tractable problem than the Vlasov--Boltzmann equation
  for the evolution of the full distribution function.

\section{Turbulent heat flux}
  With the knowledge of the distribution function, one can calculate any
  characteristics of the turbulence. In particular, the turbulent heat flux as
  a function of the temperature gradient, flow shear, or any other equilibrium
  parameters is a key quantity of interest. Focusing on the heat flux will
  allow us to diagnose the transition to turbulence, as values greater than
  zero indicate a turbulent state.  We focus here on the heat flux due to the
  ions (deuterium in our simulations). The radially outward, time-averaged ion
  heat (energy) flux through a volume $V$ enveloping a given flux surface is
  \begin{equation}
    Q_i = \left\langle\frac{1}{V}\int \mathrm{d}^3 \vb*{r} \int \mathrm{d}^3
    \vb*{v} \frac{m_i v^2}{2} \delta f_i \vb*{V\!}_E \cdot \nabla r \right\rangle,
    \label{Qdef}
  \end{equation}
  where $\vb*{V\!}_E$ is the $\vb*{E}\times\vb*{B}$ drift velocity due to the
  perturbed electric field, calculated from $\delta f_s$ via the plasma
  quasineutrality constraint (see appendix~\ref{App:gk_numerics}), and $\left<
  \ldots\right>$ indicates an average in time. $Q_i$ is typically normalised by
  the so-called gyro-Bohm value, $Q_{\rm gB} = n_i T_i v_{{\rm th}i}\rho_i^2/
  a^2$, where $n_i$, $T_i$, and $\rho_i$ are the ion density, temperature, and
  Larmor radius, respectively (it is a feature of the asymptotic ordering on
  which the gyrokinetic theory is based that $Q_i/Q_{\rm gB}$ is a finite
  number; see~\citealt{Abel2013}).

\begin{figure}
    \centering
    \includegraphics[width=0.5\textwidth]{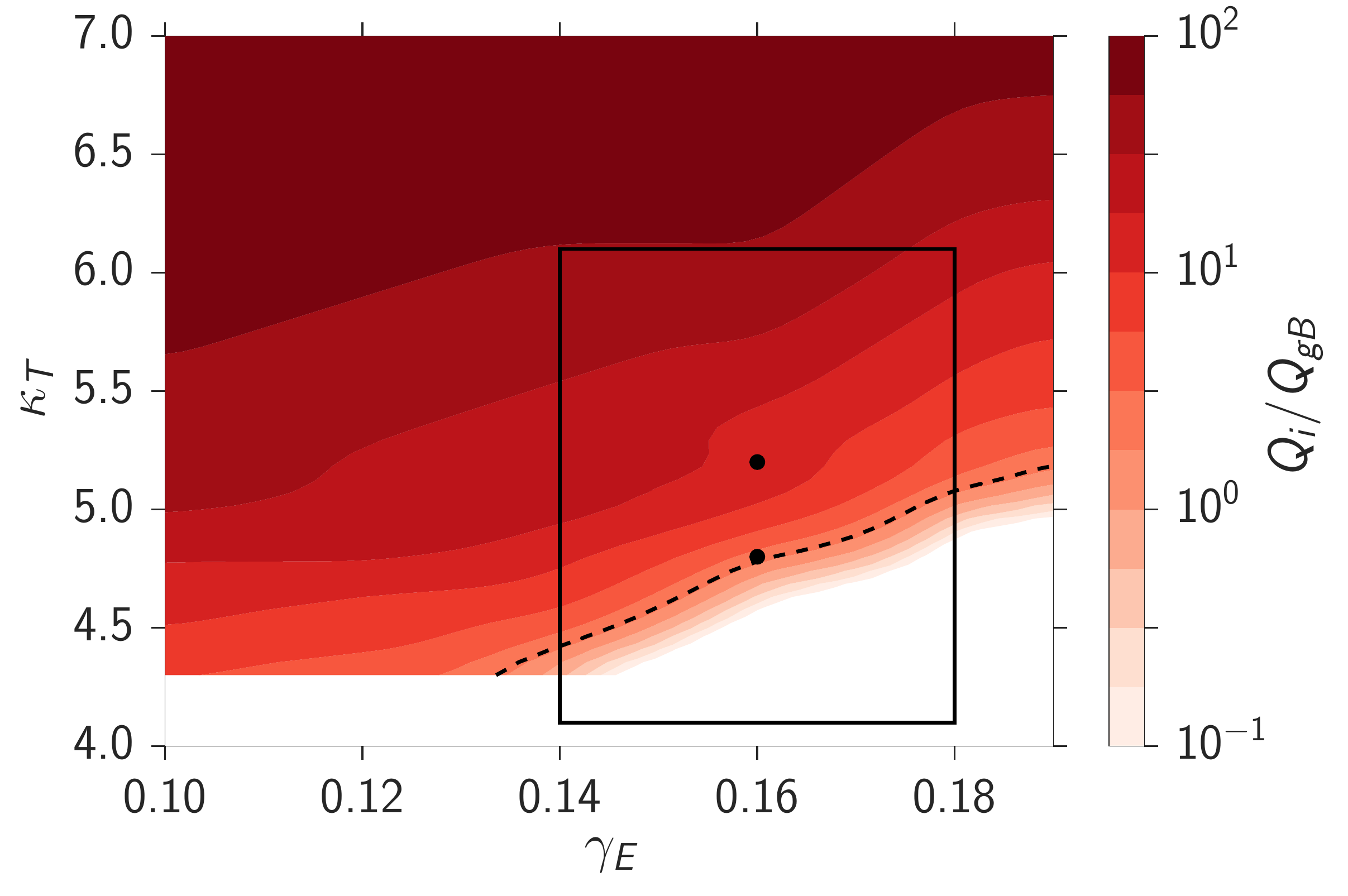}
    \caption{Normalised turbulent heat flux as a function of flow
      shear $\gamma_E$ and temperature gradient $\kappa_T$. The marked region
      indicates the range of $(\gamma_E,\kappa_T)$ compatible with the
      experimentally measured values on our chosen flux surface and the dashed
      line is the contour corresponding to the experimental value $Q_i/Q_{\rm
      gB}\approx2$~\citep{Field2014a}. Thus, the experiment is close to the
      turbulence threshold.  The two black dots mark the cases for which the
      density-fluctuation field is shown in \figref{density_fluc}.
    }
    \label{fig:heatmap}
  \end{figure}

  To map out the transition to turbulence in our system, we vary the flow shear
  $\gamma_E$ and the temperature gradient $\kappa_T$ around their experimental
  values ($\gamma_E = 0.16 \pm 0.02$ and $\kappa_T = 5.1 \pm 1$), and covering
  the range $\gamma_E \in [0,0.19]$ and $\kappa_T \in [4.3,8.0]$.
  \Figref{heatmap} shows the turbulent heat flux $Q_i/Q_{\rm gB}$ in a part of
  this range, close to the threshold. All simulations were run until they
  reached a statistically steady state, i.e.,\ until the running time average
  value became independent of time. An average was then taken over a period of
  $\sim 100 (a/v_{\mathrm{th}i})$ (which represents $\sim 400$ $\mu$s) during
  this steady state. Examining both the range of values of
  $(\gamma_E,\kappa_T)$ compatible with experiment and the experimentally
  determined value $Q_{i,\exp}/Q_{\rm gB} \approx 2.0 \equiv Q_{\rm
  \exp}$~\citep{Field2014a}, we see that {\em the turbulent state found in a
  real device is close to the threshold\/} (perhaps not a surprising
  conclusion, but an important one to be able to make quantitatively).
  Experimental investigations (e.g.,~\citealt{Mantica2009}) corroborate this
  observation for other tokamaks and show that the proximity to threshold is
  enforced by the rapid increase in $Q_i$ as the stability parameter $\kappa_T$
  is increased from its critical value (``stiff transport''). A similar
  conclusion can be drawn from~\figref{heatmap}: small increases in $\kappa_T$
  lead to order-of-magnitude changes in $Q_i$. We find that very small
  departures of the flow shear $\gamma_E$ from the threshold also lead to large
  increases of $Q_i$, showing that {\em flow shear matters at experimentally
    relevant values and heat transport is highly sensitive to it}.

\section{Subcritical turbulence}
\begin{figure}
    \centering
    \begin{subfigure}[t]{0.49\textwidth}
      \includegraphics[width=\textwidth]{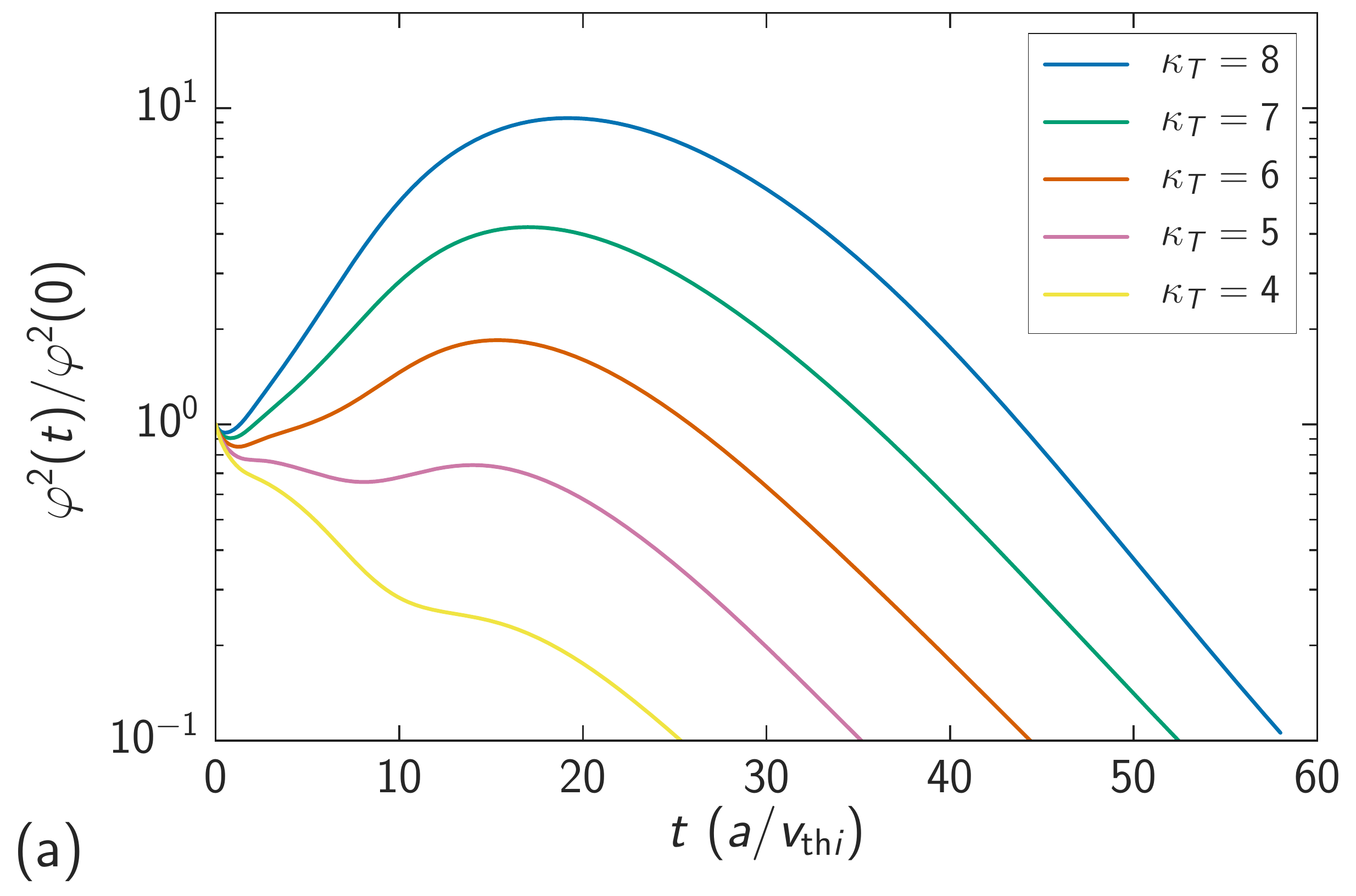}
      \caption{}
      \label{fig:transient}
    \end{subfigure}
    \hfill
    \begin{subfigure}[t]{0.49\textwidth}
      \includegraphics[width=\textwidth]{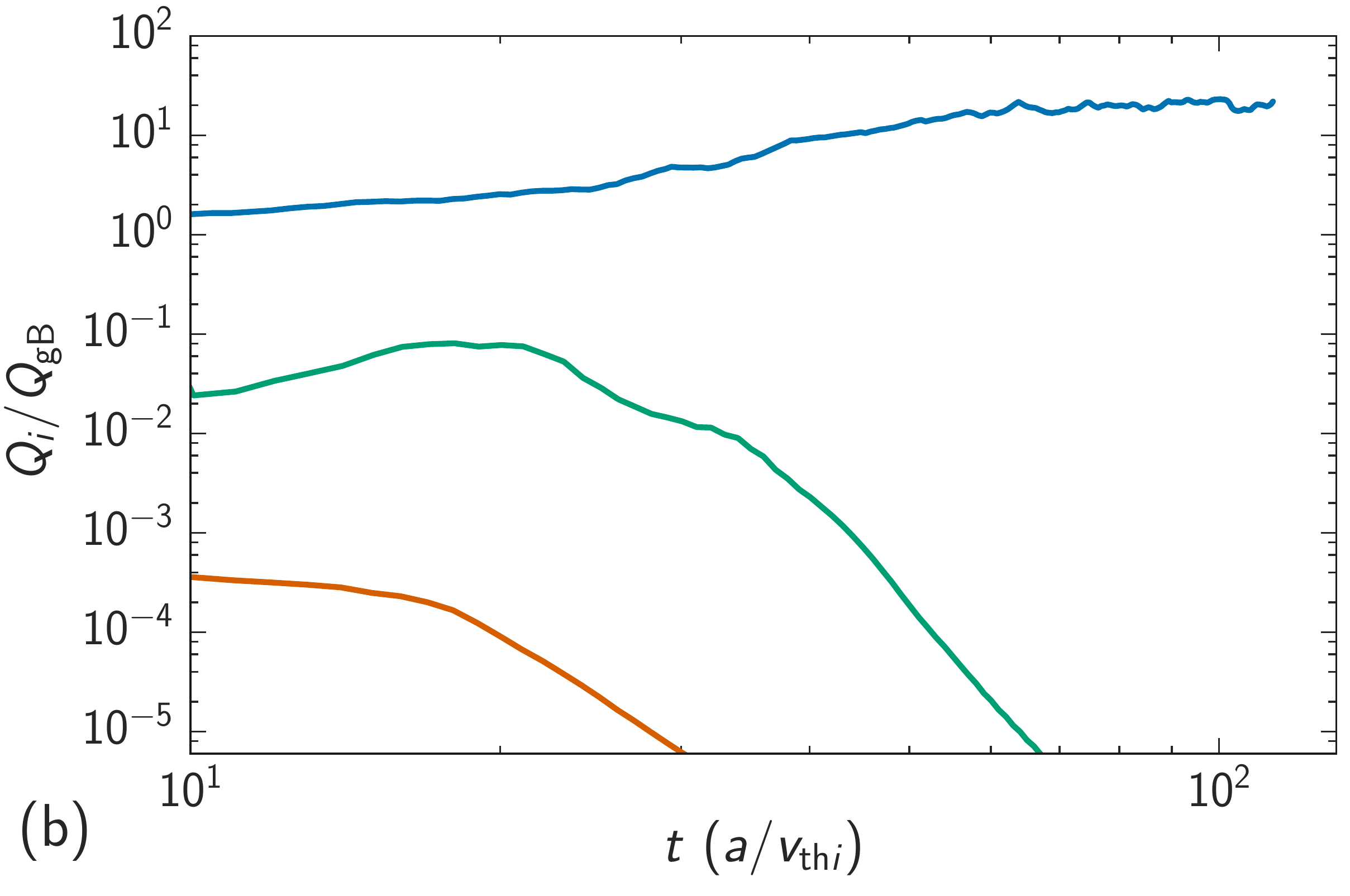}
      \caption{}
      \label{fig:saturated}
    \end{subfigure}
    \caption{(a) Transient growth of initial perturbations
      of the electrostatic potential $\varphi$ for $\gamma_E = 0.16$
      (experimentally measured value) and several values of the ion temperature
      gradient $\kappa_T$. These time-evolution curves are obtained in purely
      linear simulations, for a single mode poloidal wave number
      $k_y\rho_i=0.2$, which is approximately the wave number of the most
      vigorous transient growth. The radial wavenumber, $k_r$, is allowed to
      vary with time to resolve the effect of flow shear that causes growing
      modes to be shifted to neighbouring wavenumbers.  As the stability
      parameter $\kappa_T$ increases away from the (nonlinear) threshold (cf.\
      \figref{heatmap}), initial perturbations are amplified by an ever larger
      factor before decaying. (b) Illustration of
      subcritical turbulent state (as measured by the normalised heat flux,
      $Q_i/Q_{\rm gB}$) being reached starting from a finite perturbation,
      whereas small perturbations decay. The three nonlinear simulations shown
      here differed only in their initial perturbation level, all other
      parameters being the same and lying within the window of experimentally
      consistent values (see \figref{heatmap}), $\gamma_E=0.16$ and $\kappa_T =
      5.5$ (so the linear transient amplification in this case is very low; see
      (a)). The time histories start at $t = 10$ because we
      excluded some initial evolution involving various numerical adjustments.
    }
    \label{fig:phiiinit}
  \end{figure}

  Usually one would also carry out a sequence of linear simulations to
  ascertain whether the turbulence threshold found nonlinearly coincides with
  the linear stability boundary. Doing these simulations showed that all modes
  in our system at all parameter values that we have investigated (except
  $\gamma_E=0$) were formally linearly stable. Initial perturbations did,
  however, exhibit transient growth, typically for a longer period in the cases
  far from the nonlinear threshold, as illustrated in \figref{transient}
  (cf.~\citealt{Newton2010,Highcock2010,Barnes2011a,Schekochihin2012a,Highcock2012}).
  Nonlinearly, this means that, beyond a certain threshold in $\gamma_E$ and
  $\kappa_T$, and given a large enough initial perturbation, {\em subcritical
  turbulence\/} can be sustained.\footnote{Note in \figref{transient} that the
  turbulence threshold corresponds roughly to the values of the stability
  parameters $\gamma_E$ and $\kappa_T$ at which the transient amplification
  factor drops below unity (cf.~\citealt{Highcock2012}).} This is illustrated
  in \figref{saturated}, showing the effect of changing the initial
  perturbation amplitude. There is clearly a critical amplitude above which the
  nonlinearity can pick up the transiently amplified perturbations (very weakly
  amplified, when close to threshold) and give rise to a non-zero saturated
  state. Importantly, the saturation level does not depend on the size of the
  initial perturbation (as long as the latter is large enough). Thus, in the
  experimental instance that we have considered, {\em ion-scale turbulence in
  MAST in the presence of flow shear is subcritical\/} and so tokamak plasmas
  join a plethora of neutral fluid systems where the transition to turbulence
  depends strongly on the (size of) initial
  perturbation~\citep{Trefethen1993,Darbyshire1995}: e.g., both Poiseuille and
  Couette flows are formally stable~\citep{Salwen1980,Trefethen1993}, but still
  able to transition to a nonlinear, turbulent state; a similar situation
  arises in Keplerian shear flows believed to exist in accretion
  disks~\citep{Riols2013}. Recent theoretical work, involving very simple
  models, suggested that this may also be possible in
  plasmas~\citep{Newton2010,Schekochihin2012a,Landreman2015}, as did
  simulations of simplified tokamak
  equilibria~\citep{Barnes2011a,Highcock2010,Highcock2011,Highcock2012}, but
  ours appears to be the first demonstration of subcritical ion-scale
  turbulence in a specific, experimentally diagnosed tokamak plasma.

\section{Scenario for transition to turbulence}
  How does the transition to subcritical turbulence occur, i.e., what
  sequence of turbulent states does the system go through as either $\gamma_E$
  or $\kappa_T$ crosses the critical threshold and moves away from it into ever
  more strongly driven regimes? It is clear that the transition cannot occur
  via near-threshold states featuring arbitrarily small fluctuations
  everywhere, because sustaining subcritical turbulence requires finite initial
  perturbations. These initial perturbations must be larger near the threshold
  than far from it because the amount of amplification expected during
  transient growth tends to decrease close to the
  threshold~\citep{Schekochihin2012a,Highcock2012}. If the typical maximum
  amplitude of the fluctuations in the saturated state must remain finite, one
  way to reduce the turbulent heat flux to low values near the threshold is by
  reducing the fraction of the system's volume taken up by turbulence, i.e., by
  concentrating intense fluctuations in a shrinking part of space. This is
  precisely what happens, as we will now demonstrate.

\begin{figure}
    \centering
    \begin{subfigure}[t]{0.49\textwidth}
        \includegraphics[width=\textwidth]{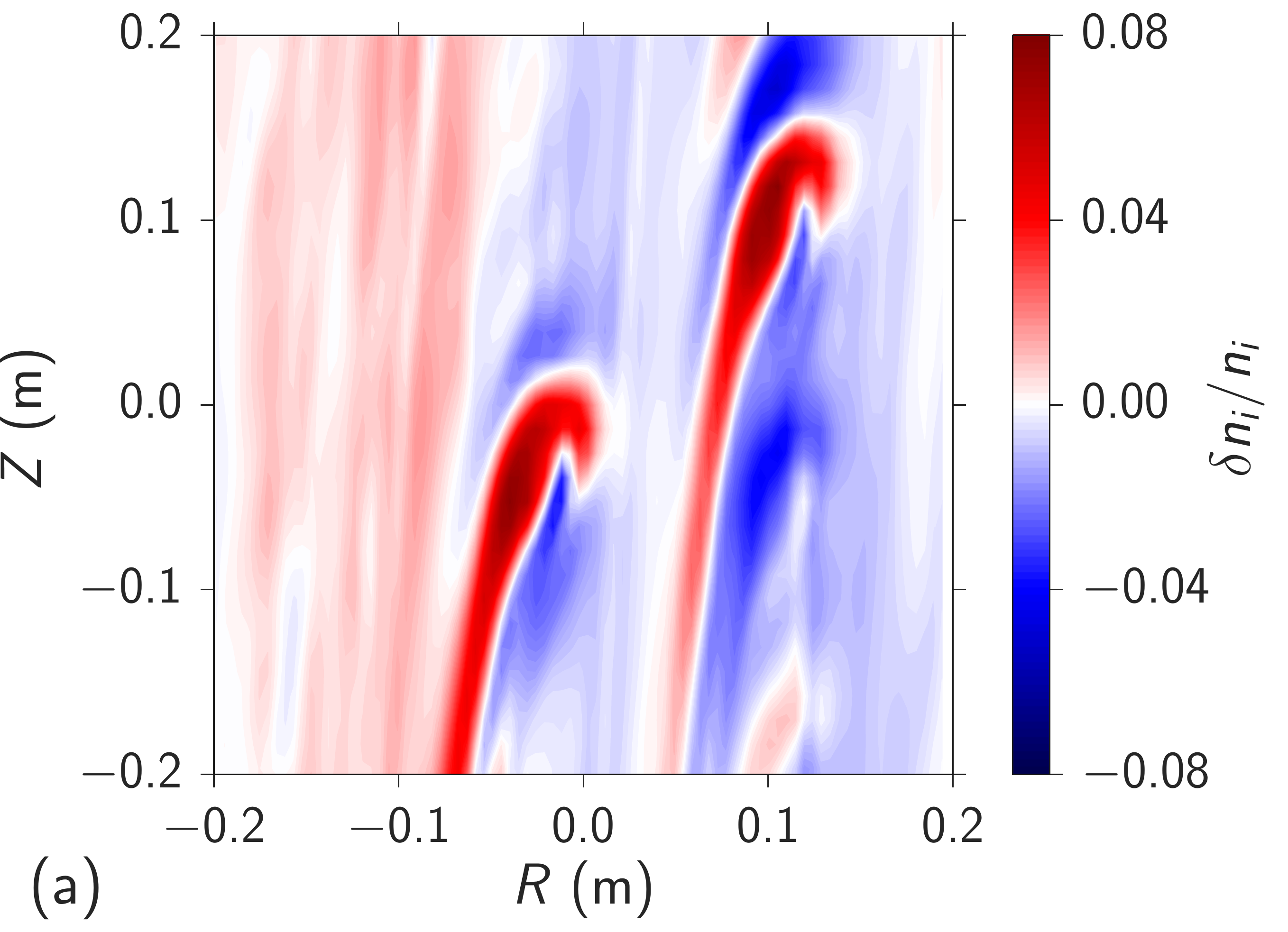}
        \caption{}
        \label{fig:marginal}
    \end{subfigure}
    \hfill
    \begin{subfigure}[t]{0.49\textwidth}
        \includegraphics[width=\textwidth]{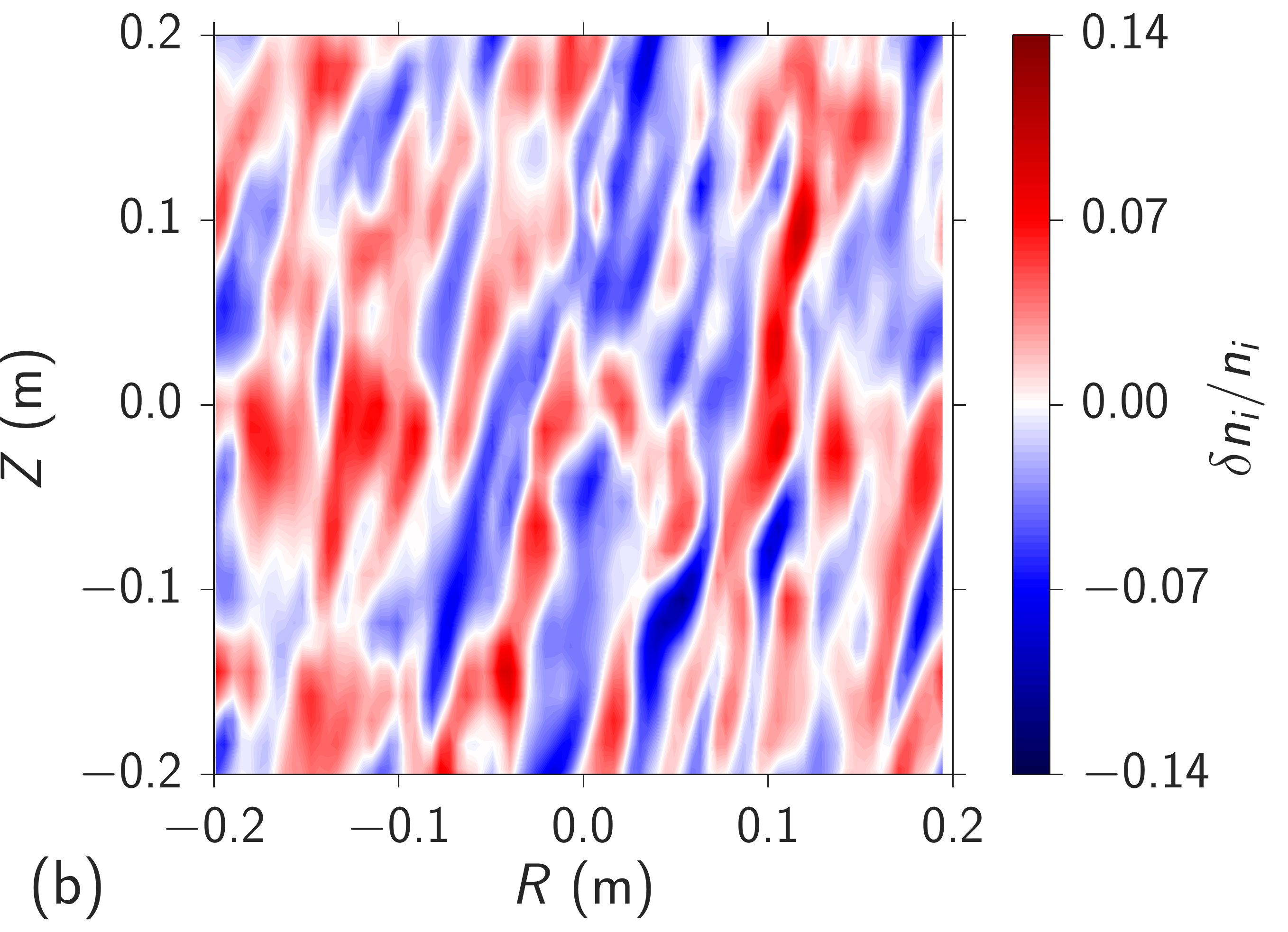}
        \caption{}
        \label{fig:strongly_driven}
    \end{subfigure}
    \caption{Density-fluctuation field $\delta n_i/n_i$ in the poloidal plane
      for simulations close to and far from the turbulence threshold (the
      two cases marked by points in \figref{heatmap}): (a) $\gamma_E = 0.16$
      and $\kappa_T = 4.8$ (near threshold), (b) $\gamma_E = 0.16$ and $\kappa_T
      = 5.2$ (strongly driven). Half of the full simulation domain in the
      vertical direction is shown.
    }
    \label{fig:density_fluc}
  \end{figure}

  In \figref{density_fluc}, we show real-space snapshots of the turbulent
  density-fluctuation field in the poloidal cross section of our flux tube at
  $\gamma_E = 0.16$ and for two different temperature gradients: $\kappa_T =
  4.8$, which is very close to the threshold, and $\kappa_T = 5.2$, a case that
  represents more strongly driven turbulence away from the threshold (both
  points are within the experimentally consistent range; see \figref{heatmap}).
  We find that the near-threshold turbulent state is dominated by
  long-lived, intense coherent structures, which travel across the domain both
  radially and poloidally, whereas far from the threshold, we observe a more
  conventional chaotic turbulent state characterised by interacting eddies.
  These two cases are representative of the relevant regions of our parameter
  space. We always find that long-lived, large-amplitude structures form in the
  near-threshold cases and survive against the background of very weak ambient
  fluctuations. As the system is taken away from the threshold by increasing
  $\kappa_T$ or decreasing $\gamma_E$, these structures become more numerous
  (i.e., more volume-filling) while retaining comparable amplitude, eventually
  start interacting with each other, and break up. Finally, far from the
  threshold, we observe no discernible long-lived structures, but rather strong
  time-variable fluctuations everywhere with amplitudes that increase with
  $\kappa_T$ or decreasing $\gamma_E$.

\begin{figure}
    \centering
    \begin{subfigure}[t]{0.49\textwidth}
      \includegraphics[width=\textwidth]{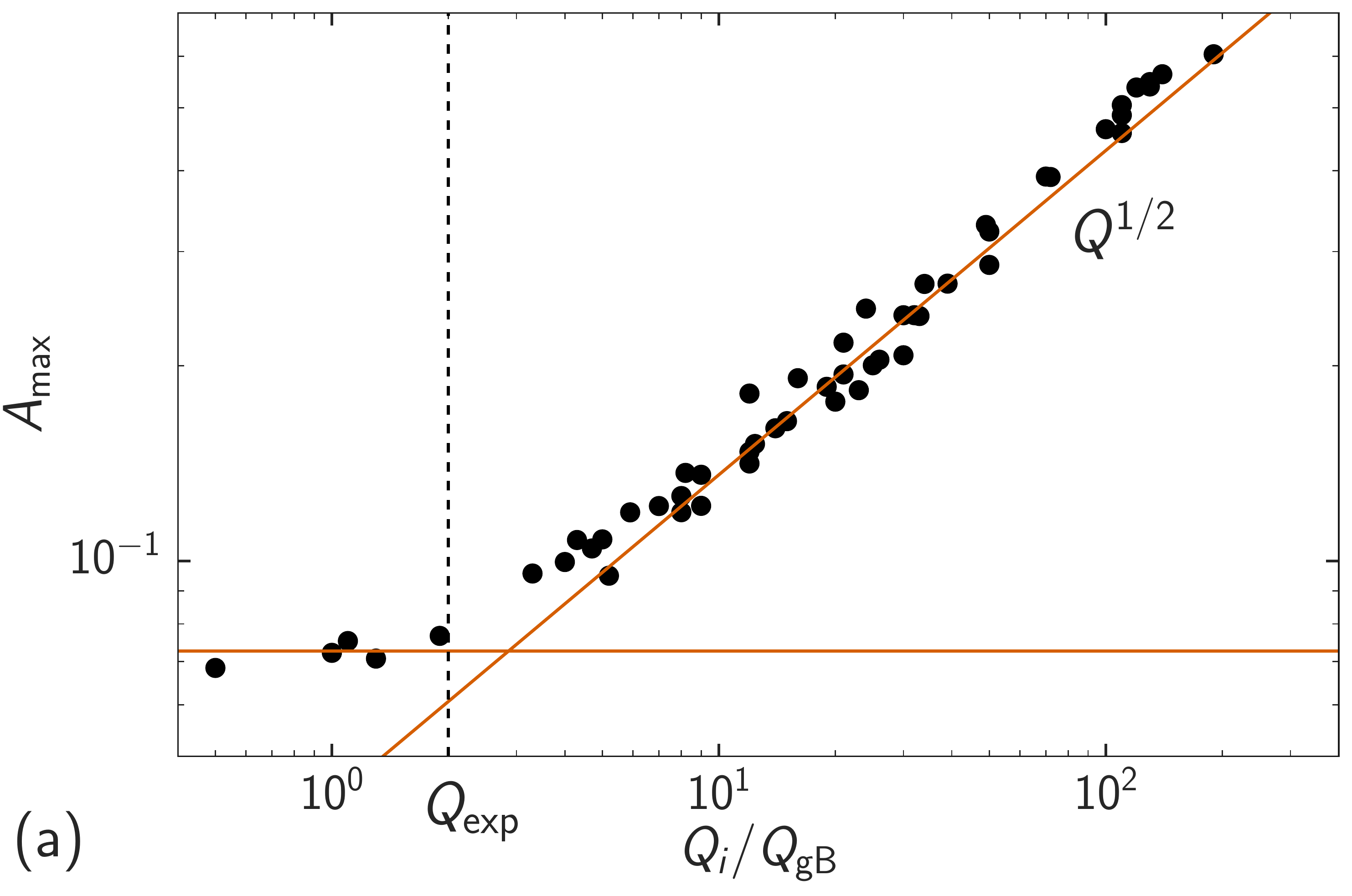}
      \caption{}
      \label{fig:amplitude}
    \end{subfigure}
    \hfill
    \begin{subfigure}[t]{0.49\textwidth}
      \includegraphics[width=\textwidth]{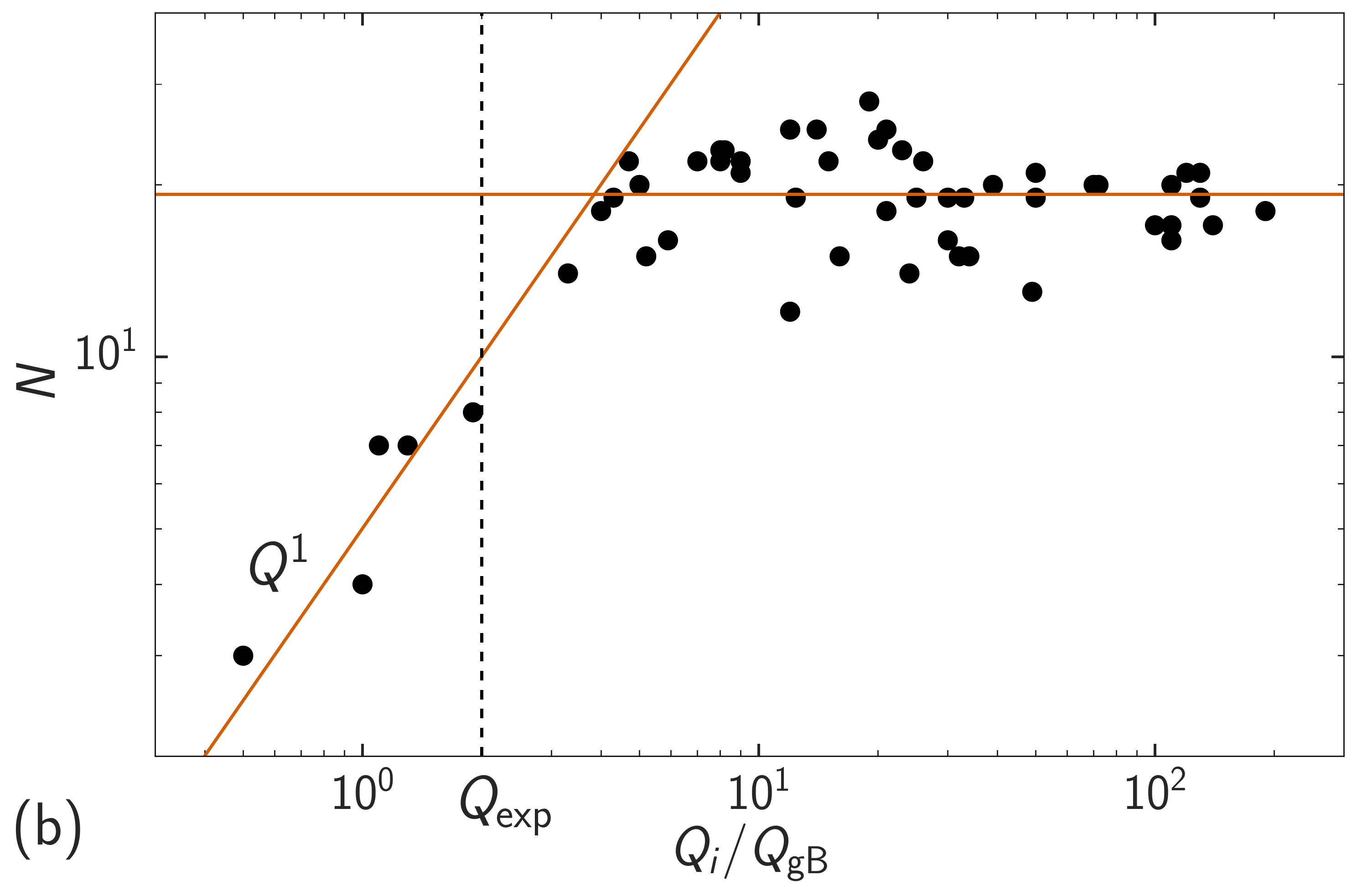}
      \caption{}
      \label{fig:nblobs}
    \end{subfigure}
    \caption{(a) Maximum amplitude of the density fluctuations versus the time
      averaged ion heat flux. The naive scaling~\eqref{q_scaling} is shown and
      holds far from threshold, whereas for small values of $Q_i/Q_{\rm gB}$
      (around and below the experimental value $Q_{\exp}$), the amplitude
      flattens. (b) Number of turbulent structures (amplitude within 75\% of
      the maximum) versus the time averaged ion heat flux. It grows up to and
      slightly beyond the experimental value $Q_{\exp}$, eventually the volume
      is filled with turbulent structures and their number tends to a constant.
      The scaling $Q\propto N$ is shown for reference: the heat flux near
      threshold is controlled by the accumulation of distinct structures; far
      from the threshold, the number of structures is simply set by their scale
      and the size of the domain, whereas the heat flux is controlled by the
      overall fluctuation amplitude [see (a)].}
  \end{figure}

  Let us make the statements above more quantitative. Consider how the {\em
  maximum amplitude\/} $A_{\rm \max}$ of the density perturbations found across
  the domain (and averaged over time) changes as our stability parameters
  $\kappa_T$ and $\gamma_E$ change. Since $Q_i$ is a strong function of both
  $\kappa_T$ and $\gamma_E$, we can measure the ``distance from threshold'' by
  just using $Q_i/Q_{\rm gB}$ as a stability parameter. A naive estimate based
  on~\eqref{Qdef} is
  \begin{equation}
    \frac{Q_i}{Q_{\rm gB}}\sim \frac{a^2}{\rho_i^2}\frac{\delta
    n_i}{n_i}\frac{V_{Er}}{v_{{\rm th}i}} \sim k_y\rho_i \frac{T_e}{T_i}
    {\left(\frac{a}{\rho_i}\frac{\delta n_i}{n_i}\right)}^2 \sim A^2,
    \label{q_scaling}
  \end{equation}
  where $A= (a/\rho_i)\delta n_i/n_i$ (which in the gyrokinetic theory is an
  order-unity quantity; see~\citealt{Abel2013}). We have estimated the radial
  $\vb*{E}\times\vb*{B}$ velocity as $V_{Er}\sim (c/B) k_y \varphi \sim
  k_y\rho_i v_{{\rm th}i} e\varphi/T_i$, where $k_y$ is the typical poloidal
  wave number ($\sim \rho_i^{-1}$ in this regime) of the fluctuations of the
  electrostatic potential $\varphi$, which are related (by order of magnitude)
  to the electron (and, therefore, ion) density via the Boltzmann response
  $e\varphi/T_e \sim \delta n_e/n_e$. \Figref{amplitude} shows the
  relationship between $A_{\rm \max}$ and $Q_i/Q_{\rm gB}$ for a number of
  simulations with different values of $\kappa_T$ and $\gamma_E$.  While the
  naive scaling~\eqref{q_scaling} is indeed manifest far from the threshold, it
  is a striking feature of \figref{amplitude} that the maximum fluctuation
  amplitude hits a finite ``floor'' as $Q_i/Q_{\rm gB}$ decreases to and below
  its experimental value --- this coincides with the appearance of long-lived
  structures illustrated in \figref{marginal}. Thus, while in the conventional
  supercritical turbulence, we might have observed smaller fluctuation
  amplitudes corresponding to lower heat fluxes all the way to the threshold,
  in the present subcritical turbulent system, we see the heat flux decrease
  while the maximum fluctuation amplitude remains constant.

  As we argued above, it does this via reduction of the volume taken up by
  large fluctuations.  We will now show this by measuring the typical number of
  the turbulent structures as a function of distance to threshold. While 2D
  structures are easily discerned by a human eye (e.g., in \figref{marginal},
  there are two), counting them systematically is a not entirely trivial
  problem, which is often encountered in computer-vision and pattern-recognition
  applications. It has been considered before in the context of experimental
  measurements of turbulence (\citealt{Kauschke1999, Muller2005}; see review of
  various techniques by~\citealt{Love2007}). Structure counting can be
  reduced to an image-labelling, or segmentation, problem by applying a
  threshold function to our density fluctuations: setting values below a
  certain percentile (here 75\%) of the maximum amplitude to 0 and above it to
  1. We are then left with an array of 1's representing our structures against
  a background of 0's. We employ a general-purpose image processing package
  \emph{scikit-image}~\citep{scikit-image}, which implements an efficient
  labelling algorithm~\citep{Fiorio1996}, to label connected regions, i.e.,
  turbulent structures. In order to improve the reliability of the labelling
  algorithm, we applied a Gaussian image filter (with a standard deviation on
  the order of the grid scale) as a pre-processing step and also removed
  structures below 10\% of the mean structure size as a post-processing step.
  These steps are justified because we are hunting intense, relatively
  large-scale structures.

  \Figref{nblobs} shows the results of the above analysis: the number $N$ of
  turbulent structures with amplitudes above the 75$^{\rm th}$ percentile
  versus the heat flux.  As in \figref{amplitude}, there are two distinct
  regimes: $N$ grows with $Q_i$ until the structures have filled the simulation
  domain (which happens just beyond the experimental value of the flux),
  whereupon $N$ tends to a constant.  Taking figures~\ref{fig:amplitude}
  and~\ref{fig:nblobs} in combination, we have, roughly, $Q_i/Q_{\rm gB}\sim N
  A^2$, i.e., {\em near the threshold, turbulent heat flux increases because
    coherent structures become more numerous (but not more intense: $N$ grows,
  $A$ stays constant), whereas far from the threshold, it does so because the
fluctuation amplitude $A$ increases (while $N$ stays constant).}

\section{Conclusion}
  In conclusion, we have discovered, using numerical simulations of an
  experimentally relevant fusion plasma, a novel scenario for transition to
  turbulent state --- which applies when the turbulence is subcritical.  Above
  a certain critical value of $\kappa_T$ and below a critical value of
  $\gamma_E$, a large enough initial perturbation will ignite turbulence. Near
  the threshold, the density and temperature fluctuations (and hence heat flux)
  are concentrated in long-lived, intense coherent structures (interestingly,
  this is reminiscent of the transition via localised patches in pipe
  flows; see~\citealt{Barkley2015}, or transition to MRI turbulence in Keplerian
  accretion flows; see~\citealt{Riols2016,Riols2013}).  As the stability
  parameters $(\kappa_T,\gamma_E)$ depart slightly from their critical values
  into the more strongly driven regime, the number of these structures
  increases rapidly while their amplitude stays roughly constant (in contrast
  to the conventional supercritical turbulence, where the amplitude increases
  with $\kappa_T$ because arbitrarily low-amplitude turbulence can be
  supported close to threshold).  Increasing the turbulent drive further leads
  to the turbulent structures filling the simulation domain and any further
  increase in the heat flux is caused by an increase in turbulent amplitude.
  The latter regime is similar to the conventional plasma turbulence.

  It is the presence of flow shear that appears to be the key feature that
  allows tokamak turbulence to exist in two distinct regimes — more strictly
  speaking, these are two regimes of anomalous transport, rather than
  turbulence: it is not obvious that the structure-dominated near-threshold
  state can be viewed as properly turbulent, representing perhaps a persistent
  nonlinear solution rather than full-scale chaos (cf. \citealt{Riols2016}). It
  will be very interesting to see if a structure-dominated regime can be
  detected in MAST or in other tokamaks where significant flow shear is
  present: so far, there are some tentative but encouraging indications that
  such a regime might manifest itself in experimentally observed skewed
  probability distributions of density fluctuations \citep{Fox2016a}.

  The new regime of tokamak turbulence described above, besides raising
  interesting questions of theory of the subcritical transition and its degree
  of universality, also raises potentially useful ones regarding ways (e.g.,
  optimal combinations of momentum and power input) in which such a turbulence
  could be controlled and the associated heat flux further reduced, leading to
  better confined plasmas.

\newpage
\section*{Acknowledgements}
  We would like to thank M.~Barnes, J.~Ball, G.~Colyer, M.~Fox and F.~Parra for
  many useful discussions.  The simulations were carried out in part using the
  HELIOS supercomputer system at International Fusion Energy Research Centre,
  Aomori, Japan, under the Broader Approach collaboration between Euratom and
  Japan, implemented by Fusion for Energy and JAEA.\@ Simulations carried out
  using the ARCHER UK National Supercomputing Service (http://www.archer.ac.uk)
  were provided by the Plasma HEC Consortium (EP/L000237/1) and the
  Collaborative Computational Project in Plasma Physics funded by UK EPSRC
  (EP/M022463/1). This work has been carried out within the framework of the
  EUROfusion Consortium and received funding from the European Union’s Horizon
  2020 research and innovation programme and Euratom research and training
  programme 2014-2018 under grant agreement number 633053.  The views and
  opinions expressed herein do not necessarily reflect those of the European
  Commission. The work of A.~A.~S. was supported in part by grants from UK
  EPSRC and STFC.

\appendix
\section{Gyrokinetic model and numerical set up}
  \label{App:gk_numerics}
  Here we outline the theoretical framework that we have used for modelling
  plasma turbulence and give all the information necessary to reproduce our GS2
  simulations.

  The gyrokinetic equation~\citep{Frieman1982, Abel2013} describes the time
  evolution of the perturbed (from a background Maxwellian $F_s$) distribution
  function $\delta f_s$ of particles of species $s$. Here we solve an
  approximate form of this equation arising from assuming, formally, that the
  Mach number $M$ of the plasma rotational flow is small, but that the flow
  shear is large enough to matter: namely, if $\omega$ is the angular frequency
  of toroidal rotation, then
  \begin{equation}
    \frac{R\omega}{v_{{\rm th}i}} = M \ll 1,\quad
    |a\nabla\ln\omega| \sim \frac{1}{M},
    \label{flow_scaling}
  \end{equation}
  where $R$ is the major radius of the toroidal device at the location of our
  flux tube. This ordering allows one to formulate local gyrokinetics in a
  rotating flux surface, neglecting such effects as Coriolis and centrifugal
  forces, but retaining flow shear. With this ordering and assuming also purely
  electrostatic perturbations (no fluctuating magnetic fields), the gyrokinetic
  system of equations is written as follows (see \S11 of~\citealt{Abel2013} or
  appendix A of~\citealt{Schekochihin2012a}).  The perturbed distribution function
  is split into a part corresponding to Boltzmann density response and the
  distribution of the gyrocentres:
  \begin{equation}
    \delta f_s = -\frac{Z_{s}e\varphi(\vb*{r})}{T_s} F_s +
    h_s(\vb*{R},\varepsilon,\mu,\sigma),
    \quad F_s = n_s(\psi){\left[\frac{m_s}{2\pi T_s(\psi)}\right]}^{3/2}
    e^{-\varepsilon/T_s(\psi)},
  \end{equation}
  where $Z_s e$, $m_s$, $n_s$, $T_s$ are the charge, mass, density, and
  temperature of particles of species $s$, $\varphi$ is the electrostatic
  potential perturbation, $\vb*{R}= \vb*{r} -
  \vb*{b}\times\vb*{v}_\perp/\Omega_s$ is the position of the centre of a
  particle's Larmor orbit, $\vb*{r}$ is the position of the particle, $\vb*{b}
  = \vb*{B}/B$ is a unit vector in the direction of the magnetic field $B$,
  $\vb*{v}_\perp$ is the velocity of the particle perpendicular to the magnetic
  field, $\Omega_s = Z_s e B / m_s c $ is the cyclotron frequency, $c$ is the
  speed of light, $\varepsilon = m_s v^2/2$ the particle energy, $\mu = m_s
  v_\perp^2/2B$ its magnetic moment and $\sigma$ the sign of its parallel
  velocity $v_\parallel = \pm {[2(\varepsilon - \mu B)/m_s]}^{1/2}$
  ($\varepsilon$, $\mu$ and $\sigma$ are the velocity-space variables used by
  the GS2 code); the velocities are taken in the frame rotating with the
  angular frequency $\omega(\psi)$, which, like $n_s(\psi)$ and $T_s(\psi)$, is
  a function of the poloidal flux $\psi$ only.  The evolution equation for
  $h_s$ is then
  \begin{equation}
    \begin{split}
      &\left(\pdv{}{t} + \vb*{u} \cdot \nabla \right) \left(h_s
      - \frac{Z_s e \ensav{\varphi}{R}}{T_s} F_s\right) +
      \left(v_\parallel \vb*{b} + \vb*{V\!}_{{\rm D}s} +
      \ensav{\vb*{V\!}_E}{R}\right)
      \cdot \nabla{h_s} - \ensav{C[h_s]}{R} \\
      &\quad=
      -\ensav{\vb*{V\!}_E}{R} \cdot \nabla r
      \left[\dv{\ln n_s}{r} + \left(\frac{\varepsilon}{T_s} -
        \frac{3}{2}\right)\dv{\ln T_s}{r}
      + \frac{m_s v_\parallel}{T_s}\frac{R B_\phi}{B}\dv{\omega}{r}\right]F_s,
      \label{gk}
    \end{split}
  \end{equation}
  where $\vb*{u} = \omega(\psi) R^2\nabla \phi$, with $\phi$ being the toroidal
  angle, is the toroidal rotation velocity,
  \begin{equation}
    \vb*{V\!}_{{\rm D}s} = \frac{c}{Z_s e B}\vb*{b}\times \left[ m_s v_\parallel^2
              \vb*{b} \cdot \nabla \vb*{b} + \mu \nabla B \right],
    \label{v_drift}
  \end{equation}
  is the magnetic drift velocity,
  \begin{equation}
    \vb*{V\!}_E = \frac{c}{B}\vb*{b}\times\nabla\varphi
  \end{equation}
  is the $\vb*{E}\times\vb*{B}$ drift velocity, $\ensav{\dots}{R}$ is an
  average over a particle orbit at constant guiding-centre position $\vb*{R}$,
  $C[h_s]$ is the linearised collision operator, $B_\phi$ is the toroidal
  component of the magnetic field.  To calculate $\varphi$ in~\eqref{gk}, the
  quasineutrality condition is used:
  \begin{equation}
    \sum_s Z_s\delta n_s = 0
    \quad\Rightarrow\quad
    \sum_s \frac{Z_s^2 e \varphi}{T_s} n_s = \sum_s Z_s \int \mathrm{d}^3 \vb*{v}
      \ensav{h_s}{r},
    \label{quasineutrality}
  \end{equation}
  where $\ensav{\dots}{r}$ means a gyroaverage at constant $\vb*{r}$.

  The last term in~\eqref{gk} represents the advection by the gyroaveraged
  $\vb*{E}\times\vb*{B}$ velocity of the Maxwellian plasma equilibrium
  distribution characterised by $n_s$, $T_s$ and $\omega$. These are functions
  only of the flux surface, conventionally labelled by the poloidal flux
  $\psi$, but here this dependence has been converted to
  the~\citet{Miller1998b} radial coordinate $r=D/2a$, where $D$ is the diameter
  of the flux surface of interest at the elevation of the magnetic axis and
  $2a$ is the diameter of the last closed flux surface (see
  \figref{nestedflux}).  Since $r$ is also a flux-surface label, $\nabla r =
  {({\rm d}\psi/{\rm d} r)}^{-1}\nabla\psi$.  The gradients that appear in the
  right-hand side of~\eqref{gk} are sources of free energy in the plasma. In
  local flux-tube calculations, these are approximated as constant parameters
  and the following definitions for them are introduced:
  \begin{equation}
    -\dv{\ln n_s}{r} = \kappa_{n s},\quad
    -\dv{\ln T_s}{r} = \kappa_{T s},\quad
    \frac{a}{v_{{\rm th}i}}\frac{r}{q}\dv{\omega}{r} = \gamma_E,
    \label{grad_defs}
  \end{equation}
  where $q(\psi)$ is the ``safety factor''. Since this quantity in a tokamak
  is, approximately, $q \sim (r/R) B/B_{\rm p}$, where $B_{\rm
  p}=|\nabla\psi|/R$ is the poloidal component of the magnetic field,
  $\gamma_E$ has the meaning of the (non-dimensionalised) part of the toroidal
  flow shear that is perpendicular to the local magnetic field.  The special
  relevance of this quantity becomes obvious if we ``unpack'' what is meant by
  $\vb*{u}\cdot\nabla$ in the left-hand side of~\eqref{gk}. Since GS2 solves
  the gyrokinetic equation locally in the vicinity of a particular flux surface
  $\psi_0$, we may expand, inside our flux tube, $\omega \approx \omega_0 + x
  B_{\rm p} R {\rm d}\omega/{\rm d} \psi$, where $x$ is the distance from the
  flux surface.  Then\footnote{This is obtained using the representation
  $\vb*{B} = B_\phi R\nabla\phi + \nabla\psi\times\nabla\phi$ of the
  axisymmetric magnetic field in a torus. Note that only $\omega$ needs to
  be expanded because we assumed in~\eqref{flow_scaling} that it changes on a
  scale smaller than $a$ (which is the scale length of change of geometrical
  and magnetic quantities such as $R$ or $B$).}
  \begin{equation}
    \vb*{u} = \omega R^2\nabla\phi
    \approx \left(\omega_0 R + x B_{\rm p} R^2 \dv{\omega}{\psi} \right)
    \left(\frac{B_\phi}{B}\vb*{b} + \frac{B_{\rm p}}{B}\vb*{e}_y\right),
  \end{equation}
  where $\vb*{e}_y = \vb*{b}\times\nabla\psi/(B_{\rm p} R)$ is the unit vector
  in the direction perpendicular to the field line but tangent to the flux
  surface. If we now go to the frame rotating with the flux surface at the rate
  $\omega_0$ and also use the fact that, in gyrokinetics, gradients parallel to
  $\vb*{b}$ are always small compared to those perpendicular to it, we find
  \begin{equation}
    \vb*{u}\cdot\nabla \approx
    x \frac{B_{\rm p}^2 R^2}{B} \dv{\omega}{\psi}\vb*{e}_y\cdot\nabla
    = \left(\frac{q R B_{\rm p}}{rB}|\nabla r|\right) x \gamma_E
      \frac{v_{{\rm th}i}}{a} \vb*{e}_y\cdot\nabla,
  \end{equation}
  with $\gamma_E$ as defined in~\eqref{grad_defs}.  The prefactor enclosed in
  the parentheses is close to unity and so $\gamma_E$ is the
  non-dimensionalised flow shear that operates on the distribution function.
  There is also free-energy injection associated with the presence of the flow
  shear, as is manifest in the presence of a term proportional to $\gamma_E$ on
  the right-hand side of~\eqref{gk}, but, at the values of $\gamma_E$ considered
  here, the (destabilising) effect of this term, while included in our
  simulations, is irrelevant in comparison with that of the ion-temperature
  gradient (see~\citealt{Schekochihin2012a} for further details on this).

  GS2 solves the GK equations~\eqref{gk} and~\eqref{quasineutrality} in a flux
  tube following the flux surface once around the torus poloidally (see
  \figref{mast} and discussion in the main text). \Figref{nestedflux} shows the
  poloidal projection of the MAST flux surfaces with the flux surface at $r =
  0.8$ marked. The marked flux surface is the location at which we solve the GK
  equation.  It can be described as located in the outer core of the device.

  Each GS2 simulation requires input of a number of constant parameters that
  define the magnetic-field geometry (e.g., elongation of the flux surface, its
  triangularity, magnetic shear, etc.), the properties of the mix of the
  participating particle species (their masses, charges, densities,
  temperatures, collisionalities, etc.) and the local thermal equilibrium
  properties of the plasma --- in particular, the gradients defined
  in~\eqref{grad_defs}. The use of the local formulation of gyrokinetics
  requires that $\rho^* \equiv \rho_i/a \ll 1$. At the radial location
  considered in this paper, $\rho_i \approx 6 \times 10^{-3}$ m and the
  minor radius $a \approx 0.6$ m, which implies $\rho^* \sim 10^{-2}$ and
  justifies the local approximation of gyrokinetics that we have used.
  Table~\ref{tab:sim_params} gives a list of the equilibrium parameters, which
  have been determined via diagnostic measurements of the MAST discharge
  \#27268 (at $t=0.25$~s). In particular, $T_i$ and $\omega$ were obtained from
  charge-exchange-recombination spectroscopy measurements~\citep{Conway2006}
  and $T_e$ and $n_e$ were obtained from a Thomson scattering
  diagnostic~\citep{Scannell2010}. The magnetic geometry in our simulations is
  described by the~\cite{Miller1998b} parametrisation and the geometric
  parameters were obtained from an EFIT reconstruction~\citep{Lao1985} of the
  equilibrium.  All these parameters were fixed at the same values in all our
  simulations, except the ion temperature gradient $\kappa_T$ and the flow
  shear $\gamma_E$ (their experimental values are $\kappa_T=5.1$ and $\gamma_E
  = 0.16$; see \figref{heatmap} for error bars on these values).

  The resolution of our simulations (with corresponding GS2 input parameters)
  was as follows: 128 radial modes (\texttt{nx}), 96 binormal modes
  (\texttt{ny}), 20 parallel grid points (\texttt{ntheta}), 16 energy grid
  points (\texttt{negrid}), and 27 pitch-angle grid points ($\texttt{ngauss} =
  8$). The box sizes were approximately $200 \rho_i$ ($\texttt{x0} = 10$ and
  $\texttt{jtwist} = 80$) and $62 \rho_i$ ($\texttt{y0} = 10$) in the radial
  and binormal directions, respectively, and $2 \pi$ in the parallel direction,
  given that GS2 uses the poloidal angle as the parallel coordinate.
  \begin{table}
    \centering
    \begin{tabular}{r c c}
      Name & GS2 variable & Value \\

      $\beta = {8\pi n_i T_i}/{B_\mathrm{ref}^2}$ & \texttt{beta} &
      0.0047 \\

      $\beta' = \pdv*{\beta}{r}$ & \texttt{beta\_prime\_input}
      & -0.12 \\

      Effective ion charge $Z_{\mathrm{eff}}
      = {\sum_i n_i Z_i^2/|\sum_i n_i Z_i|}$ & \texttt{zeff} & 1.59 \\

      Electron collisionality $\nu_e$ & \texttt{vnewk\_2} & 0.59 \\

      Electron density $n_{eN} = n_e/n_{\mathrm{ref}}$ &
      \texttt{dens\_2} & 1.00 \\

      Electron density gradient $\kappa_{ne} = - \dv*{\ln n_e}{r}$ &
      \texttt{fprim\_2} & 2.64 \\

      Electron mass $m_{eN} = m_e/m_{\mathrm{ref}}$ &
      \texttt{mass\_2} & $1 / (2 \times 1836)$ \\

      Electron temperature $T_{eN} = T_e/T_{\mathrm{ref}}$ &
      \texttt{temp\_2} & 1.09 \\

      Electron temperature gradient $\kappa_{Te} = - \dv*{\ln T_e}{r}$ &
      \texttt{tprim\_2} &  5.77\\

      Elongation $\kappa$ & \texttt{akappa} & 1.46 \\

      Elongation derivative $\kappa' = \dv*{\kappa}{r}$ &
      \texttt{akappri} & 0.45 \\

      Flow shear $\gamma_E = (r/q) \dv*{\omega}{r} (a/v_{{\rm th}i})$ &
      \texttt{g\_exb} & [0, 0.19] \\

      Ion collisionality $\nu_i$ & \texttt{vnewk\_1} & 0.02 \\

      Ion density $n_i = n_{\mathrm{ref}}$, $n_{iN} = n_i/n_{\mathrm{ref}}$ &
      \texttt{dens\_1} & 1.00 \\

      Ion density gradient $\kappa_n = - \dv*{\ln n_i}{r}$ &
      \texttt{fprim\_1} & 2.64 \\

      Ion mass $m_i = m_{\mathrm{ref}}$, $m_{iN} = m_i/m_{\mathrm{ref}}$ &
      \texttt{mass\_1} & 1.00 \\

      Ion temperature $T_i = T_{\mathrm{ref}}$, $T_{iN} = T_i/T_{\mathrm{ref}}$ &
      \texttt{temp\_1} & 1.00 \\

      Ion temperature gradient $\kappa_T = - \dv*{\ln T_i}{r}$ &
      \texttt{tprim\_1} & [4.3, 8.0] \\

      Magnetic shear $\hat{s} = r/q \dv*{q}{r}$ & \texttt{s\_hat\_input} & 4.00\\

      Magnetic field reference point $R_\mathrm{geo}$ & \texttt{r\_geo} & 1.64 \\

      Major radius $R_{0N} = R_0/a$ & \texttt{rmaj} & 1.49 \\

      Miller radial coordinate $r = {D/2a}$ &
      \texttt{rhoc} & 0.80 \\

      Safety factor $q = \pdv*{\psi_\mathrm{tor}}{\psi_{\mathrm{pol}}}$ &
      \texttt{qinp} & 2.31\\

      Shafranov Shift $1/a \dv*{R}{r}$ & \texttt{shift} & -0.31 \\

      Triangularity $\delta$ & \texttt{tri} & 0.21 \\

      Triangularity derivative $\delta' = \dv*{\delta}{r}$ &
      \texttt{tripri} & 0.46 \\
    \end{tabular}
    \caption{GS2 simulation parameters obtained from diagnostic measurements of
             the MAST discharge \#27268 and appropriately normalised.
             Here $\psi_{\mathrm{tor}} = {\left({1/2 \pi} \right)}^2 \int_0^V
             dV \vb*{B} \cdot \nabla \phi$ is the toroidal magnetic flux and
             $\psi_{\mathrm{pol}} = {\left({1/2 \pi} \right)}^2 \int_0^V
             dV \vb*{B} \cdot \nabla \theta$ is the poloidal magnetic flux.
             See http://gyrokinetics.sourceforge.net for instructions on how
             the code is run with these parameters.
            }
    \label{tab:sim_params}
  \end{table}

\clearpage
\bibliographystyle{jpp}
\bibliography{paper}

\end{document}